\documentclass[twocolumn,pre,a4paper]{revtex4}
\usepackage{amsfonts}
\usepackage{amsmath}
\usepackage{amssymb}
\usepackage{graphicx}

\begin{document}
\title{Control of Multi-level Voltage States in 
       a Hysteretic SQUID Ring-Resonator System}
\author{P. Stiffell}
\affiliation{School of Science and Technology, 
             University of Sussex, 
             Brighton, 
             Sussex, 
             BN1 9QT, 
             U.K.}
\author{M.J. Everitt}
\affiliation{School of Science and Technology, 
             University of Sussex, 
             Brighton, 
             Sussex, 
             BN1 9QT, 
             U.K.}
\author{T.D. Clark}
\email{t.d.clark@sussex.ac.uk}
\affiliation{School of Science and Technology, 
             University of Sussex, 
             Brighton, 
             Sussex, 
             BN1 9QT, 
             U.K.}
\author{J.F. Ralph}
\affiliation{Department of Electrical Engineering and Electronics, 
             University of Liverpool,
             Brownlow Hill, 
             Liverpool,
             L69 3GJ, 
             U.K.}

\begin{abstract}
  In this  paper we study  numerical solutions to  the quasi-classical
  equations of motion for  a SQUID ring-radio frequency (rf) resonator
  system in  the regime where the  ring is highly  hysteretic. In line
  with  experiment, we  show that  for a  suitable choice  of  of ring
  circuit  parameters  the  solutions  to these  equations  of  motion
  comprise sets  of levels in  the rf voltage-current dynamics  of the
  coupled system. We further demonstrate that transitions, both up and
  down,  between these  levels  can be  controlled  by voltage  pulses
  applied to the system, thus opening up the possibility of high order
  (e.g. 10 state), multi-level logic and memory.
\end{abstract}
\maketitle

\section{Introduction}

In an earlier paper~\cite{PranceWCPSRAE99} we  reported on a new phenomena generated by
the non-linear interaction  of a SQUID ring (here,  a single Josephson
weak link  enclosed by a  thick superconducting ring) with  a parallel
$LC$  resonant (tank)  circuit.  The block  diagram  for this  coupled
system is shown in
\begin{figure}[tb]
\begin{center}
\resizebox*{0.4\textwidth}{!}{\includegraphics{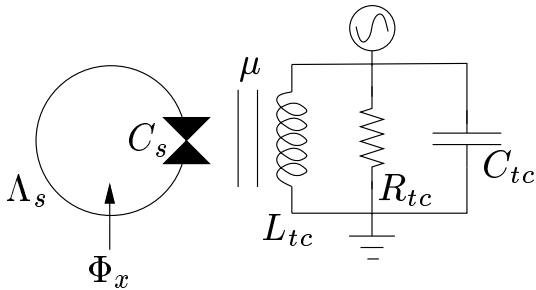}}
\caption{  Block  diagram for  an inductively  coupled SQUID  ring-tank circuit
  system  together  with  excitation,  static flux  bias  and  readout
  circuitry.  }
\label{fig:tc+s}
\end{center}
\end{figure}
figure~\ref{fig:tc+s}. As has now become apparent, SQUID rings can
display behaviour ranging from fully quantum
mechanical~\cite{FriedmanPCTL00,vanderWaltWSHOLM00,Spiller00,ChiorescuNHM03}
through to
quasi-classical~\cite{SchieveBJ88,SoerensenBCB85,Likharev96,ChuaDK87}
depending on the circuit parameters of the ring and the temperature
($T$) of the environment. At temperatures of a few K, and for
relatively large weak link capacitances $\left( 10^{-12}-10^{-13}
  \mathrm{F}\right) $, it is well
established~\cite{SchieveBJ88,Likharev96} that a SQUID ring (inductance
$\Lambda$, weak link critical current $I_{c}$) can be treated
quasi-classically, i.e. as a particle moving, with damping, in a
cosine modulated parabolic potential,
\begin{equation}
U\left(  \Phi_{s},\Phi_{x}\right)  
  =\frac{\left(  \Phi_{s}-\Phi_{x}\right)^{2}}{2\Lambda} -
   \frac{I_{c}\Phi_{0}}{2\pi}\cos\left(  \frac{2\pi\Phi_{s}}
   {\Phi_{0}}\right)
\label{squidpot}
\end{equation}
where $\Phi_{x}$  is the external  magnetic flux applied to  the ring,
$\Phi_{s}$ is the total included  flux in the ring and $\Phi_{0}\left(
  \triangleq  h/2e\right) $  is the  superconducting flux  quantum. In
this quasi-classical regime it is usual practice to describe the SQUID
ring,  and its  interactions, using  the Resistively  Shunted Junction
plus  Capacitance (RSJ+C)  model~\cite{Likharev96,BaroneP82}.  Here,  we denote  the ring
weak  link effective  capacitance  by $C_{s}$  and  the parallel  link
resistance by $R_{s}$. Furthermore, in  making use of this model it is
convenient   to  introduce   the  $\beta$-parameter   for   the  SQUID
ring $\left( \triangleq2\pi\Lambda I_{c}/\Phi_{0}\right)$. This allows
us  to distinguish  between the  parameter space  in which  $I_{c}$ is
always a  single-valued function of $\Phi_{x}$  (the inductive regime)
and  the  hysteretic  regime  where  $I_{c}$ can  be  multi-valued  in
$\Phi_{x}$~\cite{Likharev96,BaroneP82}.

The   cosine   in   the   potential  (\ref{squidpot})   leads   to   a
superconducting screening current response  in the SQUID ring which is
manifestly   a  non-linear,   and  $\Phi_{0}$-periodic,   function  of
$\Phi_{s}$.  When  at  least  part  of  this  external  flux  is  time
dependent,  and  applied   through  an  inductively  coupled  resonant
circuit,  the  ring-resonator  system  displays  non-linear  dynamical
behaviour  which has been  the focus  of much  research over  the last
three decades~\cite{SchieveBJ88,SoerensenBCB85,ClarkRPPDW98} It is precisely this behaviour that forms the
basis  of  the well  know  ac-biased  SQUID magnetometer~\cite{Likharev96,BaroneP82,Lounasmaa74}.  
Although not  invariably the case, it has  been common practice~\cite{Lounasmaa74}
to  make  use of  a  radio  frequency  (rf $\approx$  20\textrm{MHz}),
parallel $LC$  (tank circuit) resonator in  these coupled magnetometer
systems. In the work described here we adopt this frequency regime for
the  resonant circuit.  Considered on  its  own, of  course, the  tank
circuit circuit is  strictly linear, i.e. on resonance  the rf voltage
$\left( V_{out}\right) $ across it  is linearly dependent on the level
of  rf drive  input current  $\left( I_{in}\left(  t\right)  \right) $
applied.  However, when a SQUID ring is coupled to such a tank circuit
the  situation can  alter  radically. With  the  SQUID ring  potential
above,  and   at  finite  critical  current   $I_{c}$,  the  screening
supercurrent  $I_{s}=\Phi_{s}/\Lambda$ flowing in  the ring  to oppose
any  externally applied flux  $\Phi_{x}$ is  a non-linear  function of
this  flux.  Since  both the  SQUID  ring  and  the tank  circuit  are
macroscopic  in  nature,  the  back  reaction  through  the  inductive
coupling  leads to  the  system displaying  non-linear, even  chaotic,
dynamics. In  general the  stronger the non-linearity  in $I_{s}\left(
  \Phi_{x}\right) $, the more  pronounced the non-linear behaviour can
be.  This  non-linear  behaviour  is  most easily  seen  in  plots  of
$V_{out}$ versus $I_{in}\left( t\right)  $, these also being dependent
on   the  level   of  static   or  quasi-static   bias   flux  $\left(
  \Phi_{x}\right) $  applied to the SQUID ring.  In SQUID magnetometer
systems these characteristics  display essentially constant rf voltage
steps in the (time averaged) $V_{out}$ versus $I_{in}\left( t\right) $
at intervals periodic in $I_{in}$.  These SQUID steps are modulated in
a $\Phi_{0}$-periodic  manner by  the external bias  flux~\cite{Likharev96,Lounasmaa74,ZimmermanTH70}. 
The ring-tank circuit system is described dynamically by two (coupled)
equations of  motion, one for the  tank circuit and the  other for the
ring. These are given by
\begin{widetext}
\begin{alignat}{2}
 C_{tc}\frac{d^{2}\Phi_{tc}}{dt^{2}}+\frac{1}{R_{tc}}\frac{d\Phi_{tc}}
 {dt}+\frac{\Phi_{tc}}{L_{tc}\left(  1-K^{2}\right)  }
\ &=
\ I_{in}\left(  t\right) + \frac{\mu\Phi_{s}}{\Lambda\left(  1-K^{2}\right)  }
 &\qquad\text{(Tank circuit)}
\label{tcequmotion}\\
 C_{s}\frac{d^{2}\Phi_{s}}{dt^{2}}+\frac{1}{R_{s}}\frac{d\Phi_{s}}{dt}
 +I_{c}\sin\left(  \frac{2\pi\Phi_{s}}{\Phi_{0}}\right)  +\frac{\Phi_{s}
 }{\Lambda\left(  1-K^{2}\right)  }
\ &=
  \ \frac{\mu\Phi_{tc}}{\Lambda\left(1-K^{2}\right)  }
 &\qquad\text{(SQUID ring)}
\label{squidequmotion}
\end{alignat}
\end{widetext}
where the  subscripts $tc$  and $s$ refer,  respectively, to  the tank
circuit and  the SQUID ring  and $\mu\left( =M/L_{tc}\right) $  is the
fraction of  the flux coupled between  the ring and the  tank circuit. 
Thus,  $C_{tc}$  and  $L_{tc}$  are, respectively,  the  tank  circuit
capacitance  and  inductance, $\Phi_{tc}$  is  the  flux  in the  tank
circuit  inductor, $R_{tc}$  is the  resistance of  the  parallel tank
circuit on resonance  and $K\left( =\sqrt{M^{2}/\Lambda L_{tc}}\right)
$ quantifies the  strength of the inductive coupling  between the ring
and   tank   circuit   with   a   mutual   inductance   of   $M$.   In
(\ref{tcequmotion}) and  (\ref{squidequmotion}) the last  terms on the
right hand sides of these equations describe the back reaction between
the ring and the tank circuit.

In   the  original   paper~\cite{PranceWCPSRAE99}  we   used   (\ref{tcequmotion})  and
(\ref{squidequmotion})  to  compute   the  $V_{out}$  versus  $I_{in}$
dynamics  of the ring-tank  circuit system  in the  highly hysteretic,
strongly  underdamped,  regime  to  model  the  observed  experimental
behaviour.  What  was  recorded  experimentally was  a  succession  of
plateau  regions  in  the  time  averaged  $V_{out}$  versus  $I_{in}$
characteristics of the ring-tank  circuit system.  Each plateau region
consisted of a  set a parallel steps at  regular separations along the
$I_{in}$ axis.  The lengths of these  steps along this  axis were much
greater than those observed  in standard hysteretic SQUID magnetometer
characteristics, where,  typically, $\beta\approx$ a  few~\cite{Likharev96,Lounasmaa74}. In
addition,  for   each  individual  plateau  the   rf  voltage  $\left(
  V_{out}\right)   $    at   which   these    steps   occurred   varied
$\Phi_{0}$-periodically   in  $\Phi_{s}$.   In   experiment~\cite{PranceWCPSRAE99}  the
ring-tank     circuit     system      was     observed     to     jump
stochastically~\cite{GammaitoniHJM98,BulsaraG96,LindnerMNBDB01} between the  various steps associated with
each plateau. Since, in practice, it was easy experimentally to access
circuit parameters  in which  many steps per  plateau region  could be
seen (a  maximum of 19 steps/plateau  was recorded over  the course of
these experiments),  there seemed a possibility that  the plateaux and
steps could be utilised to  create multi-level logic as an alternative
to  the  standard binary  logic.  We  found  that with  simulated  low
temperature noise on the tank  circuit drive current, introduced via a
noise distribution for a thermal  bath, the solutions to these coupled
equations  of  motion modelled  our  experimental  results very  well,
including the stochastic jumping between steps on particular plateaux.
We argued in the paper~\cite{PranceWCPSRAE99}  that when the SQUID ring is sufficiently
underdamped  it  no  longer  follows  the  potential  (\ref{squidpot})
adiabatically as the rf flux  coupled in from the tank circuit changes
with  time.  Essentially,  it  is  this  non-adiabatic  response  that
generates  the  large multi-step  plateaux  in  the ring-tank  circuit
$V_{out}$ versus $I_{in}$  characteristics, each step corresponding to
a  different  flux jump  trajectory  (local  well  to local  well)  in
(\ref{squidpot}). Nevertheless,  even though we  had demonstrated that
the  underdamped  RSJ+C  description  could model  the  experimentally
observed  plateaux and  steps,  the stochastic  jumping between  steps
generated in  the theoretical calculations required  more explanation. 
In the experiments  we took the view, we  believed quite validly, that
these jumping  processes were caused  by ambient noise.  However, for
technical  reasons we  were  not  able to  measure  the full  spectral
density  function of  this  noise  and with  the  computer power  then
available to  us we could not  be certain that  the stochastic jumping
processes  seen   in  the  simulations   did  not  arise   because  of
computational   inaccuracies.   New   calculations,  at   much  higher
accuracy, are the  basis of the work reported here.  At this new level
we have been able to show  that, from the viewpoint of the simulations
presented in our previous paper~\cite{PranceWCPSRAE99}, the jumping processes arose due
to the  build up of  computational error in the  numerical integration
for  the evolution  of  the system.  Repeating  the calculations,  and
eliminating this potential problem  by using more accurate integration
methods, only single steps are  generated in each plateau region, even
with large  variance current noise  added. Nevertheless, it  was clear
from experiment  that the ring-tank circuit system  could be perturbed
to  generate  jumps.  We now  show  how  this  can  be achieved  in  a
controlled   way,  pointing   to  possible   device   applications  in
multi-level logic.

\section{SQUID ring-tank circuit dynamics in the highly hysteretic regime}
\begin{figure}[tb]
\begin{center}
\resizebox*{0.48\textwidth}{!}{\includegraphics{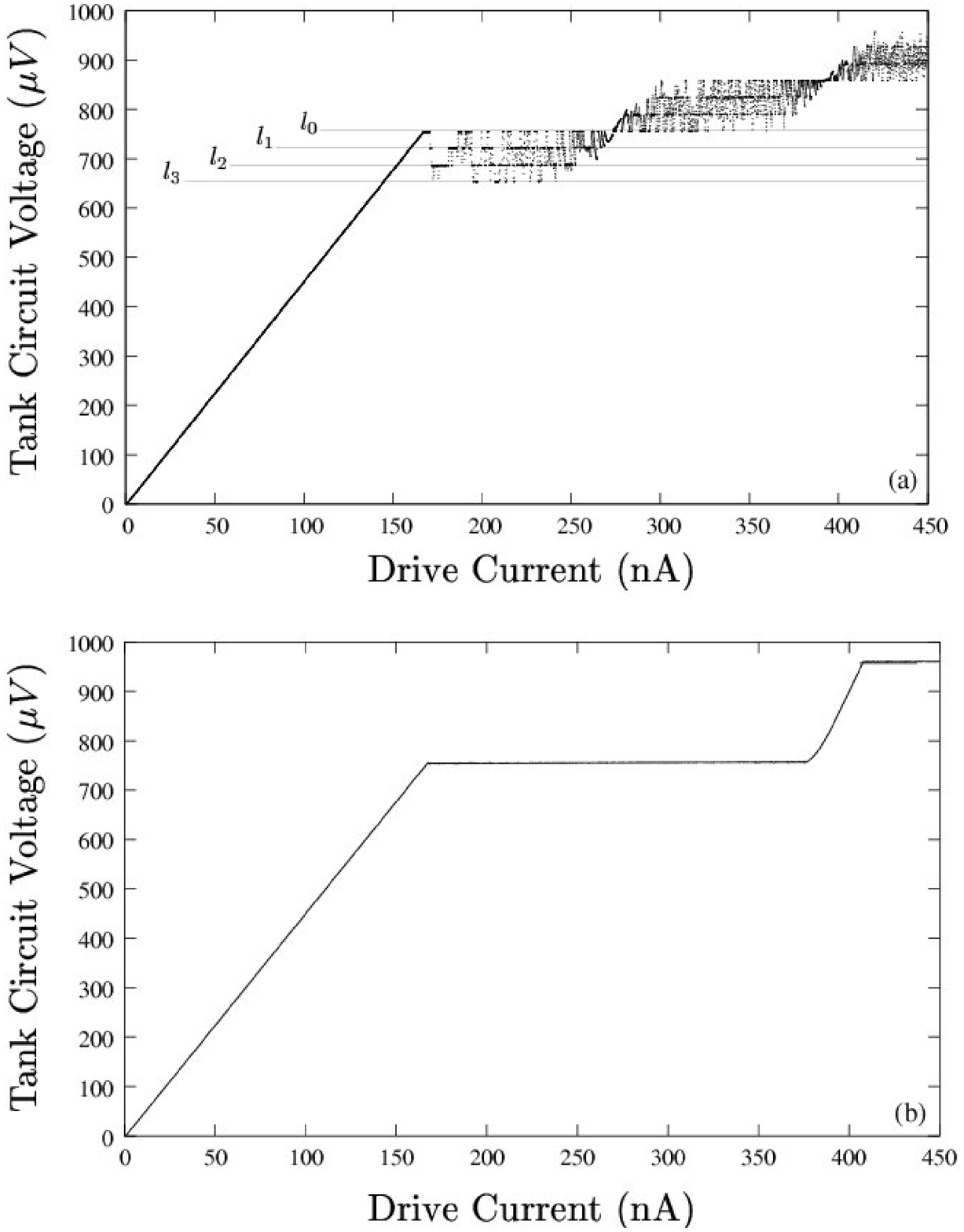}}
\caption{Dynamical rf voltage $\left(  V_{out}\right)  $ versus rf current
  $\left( I_{in}\right) $ characteristics for a highly hysteretic
  $\left( \beta=137\right) $ SQUID ring-tank circuit system with
  circuit parameters ring: $C_{sq}=1\times10^{-13}\mathrm{F}$,
  $\Lambda_{sq}=6\times10^{-10}\mathrm{H}$, $R_{sq}=10\Omega$,
  $I_{c}=75.2\mu \mathrm{A}$; tank circuit:
  $C_{tc}=7.6\times10^{-10}\mathrm{F}$, $L_{tc}=63\mathrm{nH}$ and $\mu\left(
    =M^{2}/L_{tc}\Lambda;\right) =0.0087$, a bare tank circuit
  resonant frequency $23\mathrm{MHz}$ and $\Phi_{x}=0.0\Phi_{0}$ (modulo
  $n\Phi_{0}$, $n$ integer).\vspace*{-12pt}}
\label{fig:VvI}
\end{center}
\end{figure}
\begin{figure}[b]
\begin{center}
\resizebox*{0.48\textwidth}{!}{\includegraphics{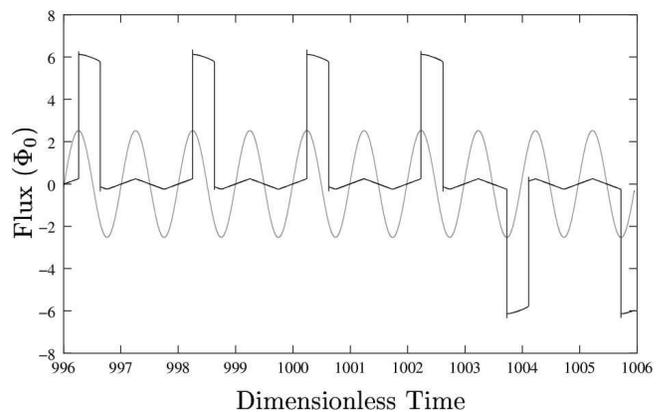}}
\caption{  Time dependence  of the  normalised SQUID ring  flux $\Phi_{sq}/\Phi
  _{0}$ as a  function of time, normalised to  the tank circuit period
  $2\pi \sqrt{L_{tc}C_{tc}}$.  }
\label{fig:SFlx}
\end{center}
\end{figure}

With access to more computational  power we considered it important to
establish  the  origin  of  the  stochastic jumping  in  the  computer
simulations  as  previously  published~\cite{PranceWCPSRAE99}.  As an  example  of  the
possible origin  of this jumping phenomenon we  computed the $V_{out}$
versus $I_{in}$  characteristics for a SQUID ring-tank  circuit in the
large $\beta$ regime. In this  we chose suitable ring and tank circuit
parameters  for   this  regime,  specifically  for   the  SQUID  ring:
$\Lambda_{s}=6\times10^{-10}$\textrm{H},      $I_{c}=75.2\mu$\textrm{A}
$\left(      \beta=137\right)$,      $C_{s}=10^{-13}$\textrm{F},
$R_{s}=10\Omega$   and  for   the   tank  circuit:   $C_{tc}=7.6\times
10^{-10}$\textrm{F}, $L_{tc}=63$\textrm{nH}  and a quality  factor $Q$
of  $500$, the  latter parameters  yielding a  bare (uncoupled  to the
ring) tank circuit resonant frequency of $23\mathrm{MHz}$. We also set
$\mu\left(  =M^{2}/L_{tc}\Lambda;\right)  =0.0087$,  a coupling  quite
typical of  low noise SQUID  ring-tank circuit systems. These  are the
circuit parameters used throughout this paper.
In figure~\ref{fig:VvI} we show  two sets of computed $V_{out}$ versus
$I_{in}$ characteristics using these ring and tank circuit parameters.
With these values of $R_{s}$ and $C_{s}$ the SQUID is underdamped. The
originally reported characteristic~\cite{PranceWCPSRAE99}, calculated at relatively low
accuracy  using  a  fourth  order  Runge-Kutta  numerical  integration
routine   with  an   adaptive  step   size  algorithm,is   plotted  in
figure~\ref{fig:VvI}(a) for  the first plateau  region. In generating
this  characteristic 4.2 Kelvin  current noise,  with a  thermal noise
distribution, has  been introduced and the  bias flux has  been set at
$\Phi_{x}=0.0\Phi_{0}$  (modulo  $n\Phi_{0}$).   Here, the  multi-step
solutions  on the first  plateau, and  the stochastic  jumping between
these,  are  perfectly clear.   Since  the  initial calculations  were
performed, we have had access  to much greater computational power and
this  has allowed  us to  apply significantly  higher accuracy  to our
numerical integration.   Subsequently we have found  that this greatly
improved accuracy  leads to the  suppression of the jumps  between the
multiple levels in the $V_{out}$ versus $I_{in}$ characteristics. This
is very apparent  in the other solution shown  in figure~\ref{fig:VvI}(b)
for which we  used exactly the same SQUID  and tank circuit parameters
as in the characteristic of figure~\ref{fig:VvI}(a). This second solution
implies that the experimentally observed  jumps may be driven by noise
processes  but not  through current  noise as  we previously  thought. 
Within the circuit model we  have adopted, the alternative is that the
driving force arises from some  form of voltage perturbation. In order
to investigate this possibility we  present a number of simulations of
the response  of this  system to voltage  pulses applied to  the SQUID
ring.
\begin{figure*}[t]
\begin{center}
\resizebox*{0.95\textwidth}{!}{\includegraphics{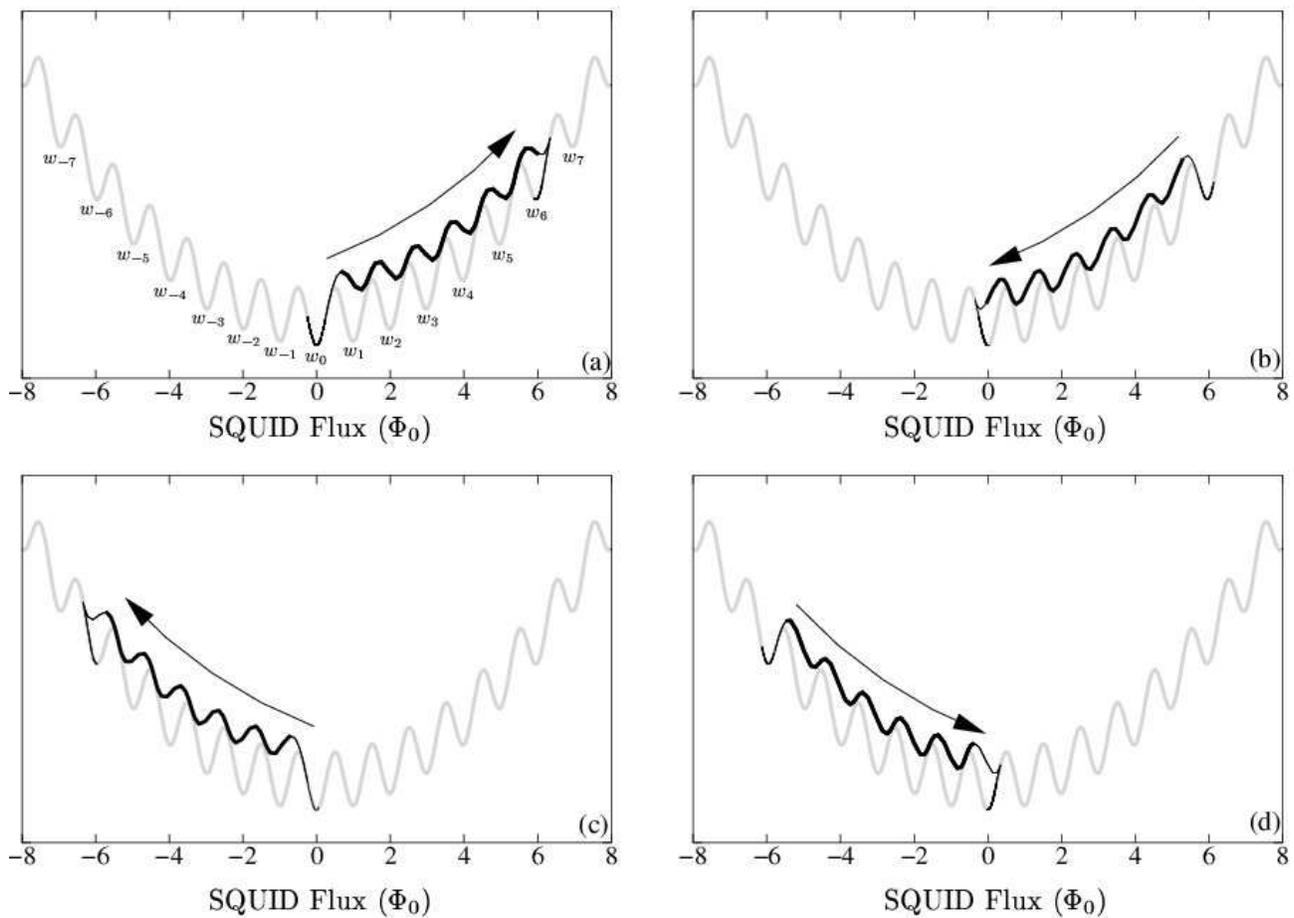}}
\caption{The SQUID ring potential for the system is shown in grey for
  \textbf{(a)} a positive  $\Phi_{s}$ going, outward, multi-$\Phi_{0}$
  trajectory; \textbf{(b)}  return trajectory for  (a); \textbf{(c)} a
  negative  going,  multi-$\Phi_{0}$  trajectory; \textbf{(d)}  return
  trajectory for (c).}
\label{fig:SFTraj}
\end{center}
\end{figure*}

In  considering the  dynamical basis  for the  plateaux (and  the step
levels) it is helpful to examine  the way in which the SQUID ring flux
$\left( \Phi _{s}\right)  $ changes in time with  the rf drive current
$\left( I_{in}\right) $. Such a response, is shown in
figure~\ref{fig:SFlx}  for a maximum  (sinusoidal) drive flux in  the tank
circuit of $13.5\Phi_{0}$ peak to  peak (equivalent to a drive flux in
the SQUID  ring of $0.1175\Phi_{0}$) .  In this figure  the SQUID ring
flux is normalised to $\Phi_{0}$ and the time axis is plotted in units
of     the     reciprocal      tank     circuit     period     $\left(
  2\pi\sqrt{L_{tc}C_{tc}}\right)$  and,   for   comparative  timing
purposes,  we show  the  form of  $I_{in}$  in light  grey. What  this
computed plot demonstrates is that, for the most part, the oscillating
flux  $\left(   \Phi_{s}\right)$  in  the   underdamped  SQUID  ring
corresponds  to  the  ring  being   confined  to  a  single  well.  In
figure~\ref{fig:SFlx} this  single well SQUID ring flux  response is shown
oscillating  over  small   amplitudes  around  $\Phi_{s}/\Phi_{0}=0$.  
However, when the drive current  reaches maximum amplitude it is found
that the SQUID ring can move outside the confines of a single well and
make  more extended  trajectories in  the  potential (\ref{squidpot}),
i.e. the flux in the ring can jump non-locally (and non-adiabatically)
several  $\Phi_{0}$  in  $\Phi_{s}$-space  between initial  and  final
(target) wells~\cite{BenjacobA81}.  We now  provide some illustrative  examples of
stable     solutions    of    equations     (\ref{tcequmotion})    and
(\ref{squidequmotion}) which give rise to the different voltage levels
seen in figure~\ref{fig:VvI}(a).

In the  particular example of figure~\ref{fig:SFlx}   these non-local jumps
take place  over $\pm$ six  wells in $\Phi_{s}$-space relative  to the
lowest energy well in the  potential. Transposed to the rf dynamics of
the coupled  system, each multi-$\Phi_{0}$  traversal leads to  one of
the  set of  plateau  step  levels in  $V_{out}$  versus $I_{in}$,  as
exemplified in  figure~\ref{fig:VvI}. Examples  of these multi-$\Phi_{0}$
trajectories (jumps) in the SQUID ring potential can be seen in
figure~\ref{fig:SFTraj}, again for the system of figure~\ref{fig:VvI}.
Taking the origin as the lowest well in this potential, the SQUID flux
trajectories  shown   in  figures~\ref{fig:SFlx}  and~\ref{fig:SFTraj}
correspond to the level denoted $l_{0}$ in figure~\ref{fig:VvI}(a). If
the origin is displaced to the  next lowest well in the potential, the
$V_{out}$ versus  $I_{in}$ characteristic of the  coupled system moves
down by  one level from the  original solution. This  level is denoted
$l_{1}$  in the solutions  plotted in  figure~\ref{fig:VvI}(a). Moving
the origin again to the next lowest well repeats the process, i.e. the
dynamics  now  correspond to  the  level  denoted  by $l_{2}$  in  the
solutions plotted in figure~\ref{fig:VvI}(a). Again, moving the origin
to  the  next lowest  well  generates  the  level denoted  $l_{3}$  in
figure~\ref{fig:VvI}(a).   However, in  our example  of  a $6\Phi_{0}$
excursion this level ($l_{3}$),  is equivalent to having displaced our
origin to  the middle well. We  find that beyond this  central well in
the excursion, attempting to localise our trajectory around one of the
remaining three wells simply reverses  the shift between the levels on
the plateau in figure~\ref{fig:VvI}(a),  i.e.  starting at $l_{3}$ and
returning to $l_{0}$ through levels $l_{2}$ and $l_{1}$. To illuminate
this discussion, we can examine,  for level $l_{0}$, the excursions in
the potential  shown in the figure~\ref{fig:SFTraj}.   Starting in the
absolute minimum well of the potential at $\Phi_{s}/\Phi _{0}=0$, with
$\Phi_{x}=0$ (modulo  $n\Phi_{0}$), the ring can  either execute local
motion  in this  well or  occasionally  move several  $\Phi_{0}$ to  a
target  well.   With the  ring  potential  plotted  in light  grey  in
figure~\ref{fig:SFTraj},     we     show     by     computation     in
figure~\ref{fig:SFTraj}(a) one  such extended trajectory,  starting in
the  lowest  well and  followed  by  a  positive going  trajectory  in
$\Phi_{s}$.  In this example, having  reached the target well the ring
completes  almost half  a  tank  circuit period  in  this well  before
returning to  its original starting  point in $\Phi_{s}$, as  shown in
figure~\ref{fig:SFTraj}(b).   This  mirrors  the  dynamical  behaviour
shown in  the first four  jumps of figure~\ref{fig:SFlx}.  It  is also
important to note  that these non-local excursions do  not occur every
time the rf drive flux  reaches its maximum amplitude, although as the
amplitude of the rf drive  grows this process happens more frequently. 
Further multi-$\Phi_{0}$ excursions in the ring potential, but now for
negative going $\Phi_{s}$,  can be seen in figures~\ref{fig:SFTraj}(c)
and~(d)  for  the  outward  and  return  paths,  respectively.   These
correspond to the last two pulses in figure~\ref{fig:SFlx}. In general the
target  well in  the  SQUID ring  is  not symmetric  about its  centre
(minimum) and this is seen to  affect the lifetime of the ring in this
target well, as  evidenced in plots presented in  figure~\ref{fig:SFlx}. In essence
the consequence  of this is  that the ring  can reach a  new traversal
(target) point in drive flux before it has completed half a $\Phi_{0}$
period in $\Phi_{s}$.  This  phenomenon, arising from the localisation
in a  non-symmetric well in the  SQUID ring potential,  means that the
excursions in $\Phi_{s}$ are only made in one direction.

\begin{figure}[tb]
\begin{center}
\resizebox*{0.48\textwidth}{!}{\includegraphics{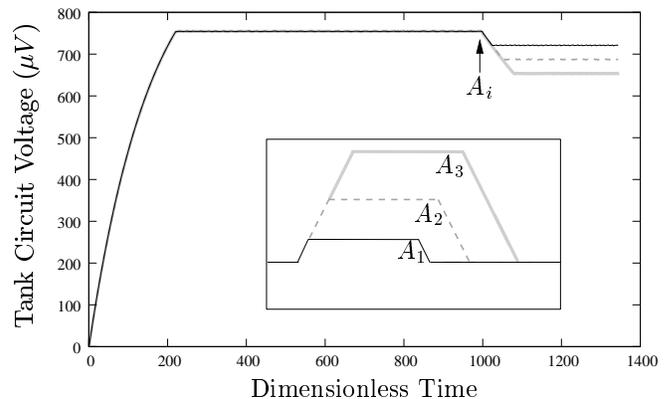}}
\caption{$V_{out}$ versus normalised time characteristics for the SQUID
  ring-tank circuit system of figure~\ref{fig:VvI} showing the controlled step to
  step jumping which can be induced using suitably configured positive
  (A-type)  ramp pulses in  the absence  of extraneous  voltage noise;
  here we show  the effect of increasing the  ramp amplitude to induce
  jumps over one, two or three step intervals.}
\label{fig:MLJ}
\end{center}
\end{figure}

With the details of figure~\ref{fig:SFTraj} in mind, and as a possible means
to    navigate    the    solutions    of    (\ref{tcequmotion})    and
(\ref{squidequmotion}  ),  we  now  consider  the  effect  of  rapidly
changing  voltages applied  to  the  SQUID ring.  In  order make  this
relatively  simple,  and   physically  transparent,  we  simulate  the
application of appropriately shaped  voltage pulses to the SQUID ring,
and  follow the  tank  circuit voltage  ($V_{out}$)  response. In  our
example we  shall now consider positive  amplitude trapezoidal voltage
pulses (inset in  figure~\ref{fig:MLJ}) with an upper  voltage state time
duration of  0.01 tank circuit  periods which is activated  during the
flux traversal shown in bold  in figure~\ref{fig:SFTraj}(d). We found that it
was only  within such highlighted  activation regions that  the system
could be made  to respond to these voltage pulses,  leading to a level
change  in  $V_{out}$. For  reference  we  term  pulses with  positive
amplitude as  A type pulse and, conversely,  negative amplitude pulses
are denoted as B type.
In figure~\ref{fig:MLJ} we have plotted the computed tank circuit voltage
response as a function of  normalised tank circuit time (i.e. in terms
of   the   tank  circuit   period)   for   the   rf  drive   amplitude
($I_{in}$=222nA) set in the middle  of the first plateau for the level
denoted $l_{0}$ in figure~\ref{fig:VvI}(a). In this calculation there was
no extraneous flux or voltage noise present and the system was allowed
to reach a steady state within  this plateau (we note that the initial
sharp  rise  at  the beginning  of  the  voltage  dynamics is  due  to
transient behaviour).  However, after a  set time, denoted  $A_{i}$ in
figure~\ref{fig:MLJ},  a voltage  pulse of the  form shown inset  in this
figure was applied. Here, the amplitudes of the pulses $A_{i},i=1,2,3$
used were  15, 75 and  120$\mu$V, respectively. As is  demonstrated in
figure~\ref{fig:MLJ}, the  application  of a  single  voltage pulse  can
change the  voltage level  of our system  by one  or more levels  at a
time, as summarised in table~\ref{tbl:ranges}.
\begin{table}[th]
 \begin{tabular}{|c|c|c|c|} \hline
Pulse Type & Lower Bound & Upper Bound & Value Used \\ \hline 
$A_1$      & 15          & 55          & 15  \\
$A_2$      & 56          & 90          & 75  \\
$A_3$      & 91          & 126         & 120 \\ \hline
 \end{tabular}
\caption{Summary of pulse amplitude ranges ($\mu$V) as described in the text.
\label{tbl:ranges}}
\end{table}
If  the  noiseless (or,  in  practice,  a  low enough  voltage  noise)
situation can be realised experimentally, the computational results of
figure~\ref{fig:MLJ} indicate that multi-level logic, based on SQUID
ring-resonator systems could  be a feasible proposition. If  so, it is
reasonable to assume  that other forms of voltage  pulse could be used
for  this purposes.   With regard  to our  example, we  note  that the
number of levels on the plateaux depends on the ring parameter values.
In the  example of  figure~\ref{fig:MLJ} this is  four but  could be
significantly larger.   To further illustrate the  control possible we
show in
\begin{figure}[tb]
\begin{center}
\resizebox*{0.48\textwidth}{!}{\includegraphics{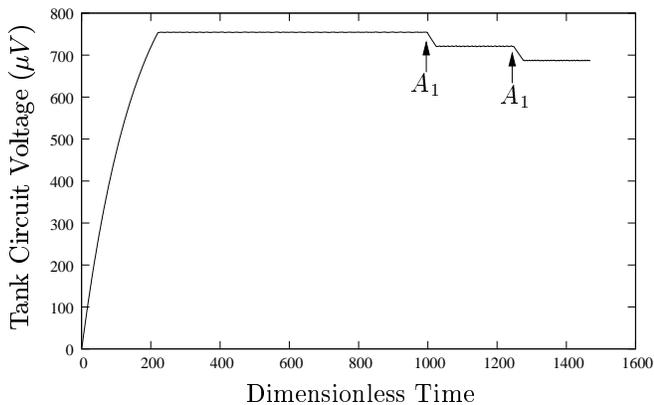}}
\caption{$V_{out}$ versus normalised time characteristics for the SQUID
  ring-tank circuit system of figure~\ref{fig:VvI} showing the effect
  of two sequential A-type voltage pulses in the absence of extraneous
  voltage noise; as is apparent these step jumps can be induced in a
  controlled manner.}
\label{fig:AAJ}
\end{center}
\end{figure}
figure~\ref{fig:AAJ}  the result of  applying two A-type pulses,  in time
sequence, to the SQUID ring-tank circuit system.

Of course, the activation pulse in voltage can be negative (B-type) as
well as  positive. The effect  of using such  a pulse after  an A-type
pulse is shown in
\begin{figure}[tb]
\begin{center}
\resizebox*{0.48\textwidth}{!}{\includegraphics{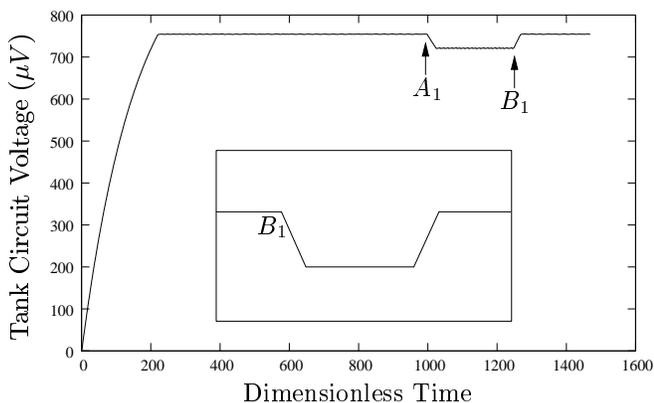}}
\caption{$V_{out}$ versus normalised time characteristics for the SQUID
  ring-tank circuit  system of figure~\ref{fig:VvI} in the absence  of extraneous
  voltage noise showing the effect of applying sequential A and B-type
  voltage ramps.}
\label{fig:ABJ}
\end{center}
\end{figure}
figure~\ref{fig:ABJ} with  the form of the B-type  pulse shown explicitly
in  the inset  of  this figure.  This  shows that  in  the absence  of
extraneous noise, and  by a suitable choice of  voltage pulse form, we
can  move at  will  between the  steps  on any  particular plateau  in
$V_{out}$ versus $I_{in}$.  To demonstrate that we can  induce step to
step jumping at essentially any time of our choosing, we show in
\begin{figure}[tb]
\begin{center}
\resizebox*{0.48\textwidth}{!}{\includegraphics{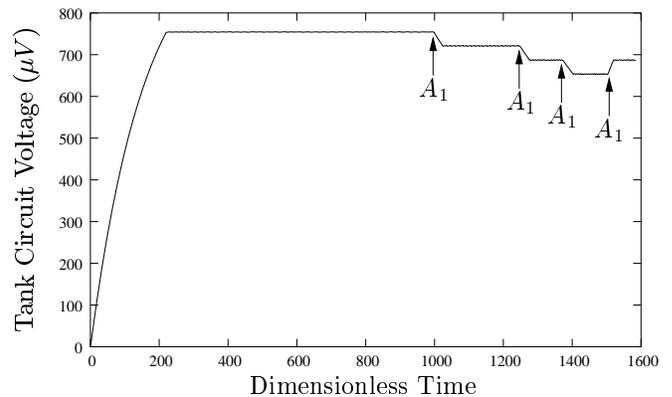}}
\caption{$V_{out}$ versus normalised time characteristics for the SQUID
  ring-tank circuit  system of figure~\ref{fig:VvI} in  the absence of
  extraneous voltage noise showing the effect of applying sequentially
  at differing time intervals as a set A-type voltage ramps.}
\label{fig:AAAAJ}
\end{center}
\end{figure}
figure~\ref{fig:AAAAJ}  a sequence of  jumps induced by a sequence of A-type
pulses  spaced   by  different  time  intervals.  We   note  that  the
application of the last pulse  causes the tank circuit voltage to step
up rather than down.

The  computed  solutions  of figures~\ref{fig:MLJ}  to~\ref{fig:AAAAJ}
show very  well the  level to level  jumping induced by  various pulse
sequences  applied  to  a  highly  hysteretic  (large  $\beta$)  SQUID
ring-tank circuit  system.  These solutions, which are  the end result
of very non-linear interactions between the ring and the tank circuit,
may be appreciated more clearly by the following qualitative argument.
As  regards  the jumping  process  between  levels  on any  particular
plateau in  $V_{out}$ versus $I_{in}$, the  system operates cyclically
from an initial  (local) well in the SQUID  potential (for example the
$w_{0}$ well  in the  potential of figure~\ref{fig:SFTraj}). This  is shown
diagrammatically in
\begin{figure}[tb]
\begin{center}
\resizebox*{0.48\textwidth}{!}{\includegraphics{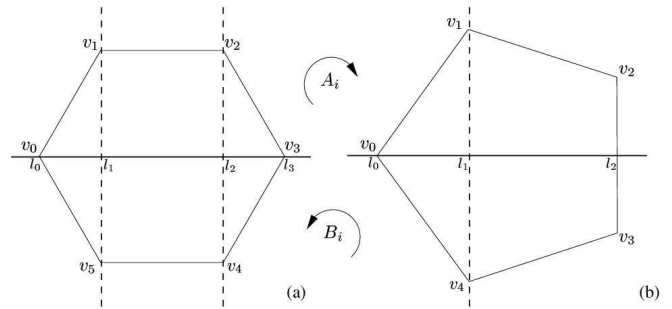}}
\caption{Graphical representation in the manner in which the wells in the
  SQUID  potential  correspond to  voltage  levels  in the  $V_{out}$
  versus  $I_{in}  $ characteristic  of  a  highly hysteretic  SQUID
  ring-tank circuit system. The  vertices around the polygons indicate
  which  well  (in a  six  well  example -  see  figure~\ref{fig:SFTraj}) the  SQUID
  oscillation is  localised in, whilst  the letters on  the projection
  denote  which  one of the set of levels on  a particular plateau in
  $V_{out}$ versus $I_{in}$ the  coupled system occupies. See text
  for more details.}
\label{fig:GrphRep}
\end{center}
\end{figure}
figure~\ref{fig:GrphRep}. The  way in  which movement  between wells  in the
SQUID potential relates  to a given voltage level  within a particular
plateau can be determined by thinking  of each well being mapped on to
one of  the vertices of a polygon  (where the order of  the polygon is
dependent  on the  parameter values  of the  ring-tank  circuit system
under investigation). This is  illustrated in figure~\ref{fig:GrphRep}(a) for
the system considered throughout this paper with the vertices labelled
$v_{i}$ for  $i=0,...,5$ and where  the mapping is  applied cyclically
from the  lowest (reference) well of  the potential on  to each vertex
(for  our  example   using  the  labels  for  the   wells  $w_{i}$  in
figure~\ref{fig:SFTraj}(a)   we   have   the   mapping   $w_{i}\rightarrow
v_{i\operatorname{mod}6}$ for  $i \in \mathbb{Z}$). Here,  we can find
the particular tank circuit  voltage level associated with each vertex
by projection  on to the  horizontal axis.  For example,  the vertices
$v_{1}$and $v_{5}$  correspond to  having the SQUID  ring-tank circuit
system in  level $l_{1}$ in figure~\ref{fig:VvI}.  Clearly, therefore, if
the system  is moved  through an appropriate  number of wells  it will
return to the same state of operation in which it began.

Using  this  graphical  map  as  the  guide, in  this  paper  we  have
investigated    a   system   with    4   levels,    corresponding   to
figure~\ref{fig:GrphRep}(a). We note that in this example the system
must  traverse 6  wells  in the  potential  before it  returns to  its
original state of operation.   If we choose different SQUID parameters
the system can  access a variable number of wells  within its cycle of
operation. However,  the system will  always display a  symmetry about
the    horizontal    axis.      For    comparison,    we    show    in
figure~\ref{fig:GrphRep}(b)  a  diagram  illustrating the  situation
where the ring-tank circuit system traverses an odd number of wells in
the SQUID potential, in this case five.

By the application of suitable  voltage pulses the system can be moved
around the vertices of the polygon, so translating the SQUID ring from
one  potential well to  another. Thus,  in figure~\ref{fig:GrphRep}(a)
application of positive (A-type) pulses causes the ring to move around
the polygon in a  clockwise direction, whilst negative (B-type) pulses
lead to anti-clockwise movement.  This movement around the polygon can
be performed utilising nearest neighbour, single well, translations, or
by  multi-$\Phi_{0}$   traversals  between  non-local   wells  in  the
potential, depending on the magnitude of the applied pulse The actual
level in  $V_{out}$ versus $I_{in}$  into which the system  settles is
determined by three  factors: (i) the well in which  the SQUID ring is
currently  localised (ii)  the size  of voltage  pulse used  to change
location in the potential and  (iii) the direction of traversal of the
polygon (whether A or B-type pulses are utilised).

\section{Conclusions}

In this work  we have demonstrated that, in  principle, it is possible
to  use  highly hysteretic  (large  $\beta$)  SQUID ring-tank  circuit
systems   as    the   basis   for   multi-level    logic   or   memory
devices~\cite{BraytonHS90,DevadasMNS88,BartlettCDH86,LinD95,LiRGWWDW98}.  Following earlier  experiments, where  giant SQUID
magnetometer  plateaux  were  observed  containing  sets  of  constant
voltage $\left( V_{out}\right)$ steps, we have shown that the jumping
processes between  steps on a  particular plateau can be  generated by
voltage pulses applied to the system.  In the absence of noise (in our
computational modelling achieved by the use of high accuracy numerical
integration techniques),  we find that a high  $\beta$ SQUID ring-tank
circuit system can remain stably on one of the steps on any particular
plateau until a suitably shaped voltage  pulse is applied to the ring. 
Following such a  pulse, the ring can be made to  jump in a controlled
manner to other steps at different levels of $V_{out}$. This new level
can be either above or  below the original voltage level, depending on
the present state  of the system and the type  of pulse used. Provided
it is possible  to reduce the ambient voltage noise  on the SQUID ring
sufficiently,  for   example  by  the  use   of  cryogenically  cooled
GaAsFET~\cite{PranceWCPSRAE99}  or   HEMT-based  pre-amplifier  electronics~\cite{BoutezCDCFB97},  it
should  prove possible to  develop devices  with a  controlled voltage
response  which can  be selected,  and modified,  at will.  This could
prove  useful in  such  areas  as multi-level  logic  or finite  state
machinery~\cite{BraytonHS90,DevadasMNS88,BartlettCDH86,LinD95,LiRGWWDW98}. As part of  any such developments we would expect
to see  interesting new fields open  up in the  non-linear dynamics of
these highly hysteretic systems.

\section{Acknowledgements}

We would like to thank  the Engineering and Physical Sciences Research
Council for its generous support of  this work. We would also to thank
Dr.  R.J.  Prance  and  Professor  A.R. Bulsara  for  interesting  and
informative discussions.

\bibliographystyle{apsrev}
\bibliography{ref}

\end{document}